\documentclass[12pt]{article}
\usepackage{epsfig}

\newcommand{\agt}{\,\rlap{\lower 3.5 pt \hbox{$\mathchar \sim$}} \raise 1pt
 \hbox {$>$}\,}
\newcommand{\alt}{\,\rlap{\lower 3.5 pt \hbox{$\mathchar \sim$}} \raise 1pt
 \hbox {$<$}\,}

\catcode`@=11
\newcount\@tempcntc
\def\@citex[#1]#2{\if@filesw\immediate\write\@auxout{\string\citation{#2}}\fi
  \@tempcnta\z@\@tempcntb\m@ne\def\@citea{}\@cite{\@for\@citeb:=#2\do
    {\@ifundefined
       {b@\@citeb}{\@citeo\@tempcntb\m@ne\@citea\def\@citea{,}{\bf ?}\@warning
       {Citation `\@citeb' on page \thepage \space undefined}}%
    {\setbox\z@\hbox{\global\@tempcntc0\csname b@\@citeb\endcsname\relax}%
     \ifnum\@tempcntc=\z@ \@citeo\@tempcntb\m@ne
       \@citea\def\@citea{,}\hbox{\csname b@\@citeb\endcsname}%
     \else
      \advance\@tempcntb\@ne
      \ifnum\@tempcntb=\@tempcntc
      \else\advance\@tempcntb\m@ne\@citeo
      \@tempcnta\@tempcntc\@tempcntb\@tempcntc\fi\fi}}\@citeo}{#1}}
\def\@citeo{\ifnum\@tempcnta>\@tempcntb\else\@citea\def\@citea{,}%
  \ifnum\@tempcnta=\@tempcntb\the\@tempcnta\else
   {\advance\@tempcnta\@ne\ifnum\@tempcnta=\@tempcntb \else \def\@citea{--}\fi
    \advance\@tempcnta\m@ne\the\@tempcnta\@citea\the\@tempcntb}\fi\fi}
\catcode`@=12
\begin{document}

\title{
\vskip-3cm{\baselineskip14pt
\centerline{\normalsize DESY 06-136\hfill ISSN 0418-9833}
\centerline{\normalsize hep-ph/0608245\hfill}
\centerline{\normalsize July 2006\hfill}}
\vskip1.5cm
\bf Prompt {\boldmath$J/\psi$\unboldmath} plus photon associated
electroproduction at DESY HERA}

\author{{\sc Bernd A. Kniehl}\\
{\normalsize II. Institut f\"ur Theoretische Physik, Universit\"at Hamburg,}\\
{\normalsize Luruper Chaussee 149, 22761 Hamburg, Germany}\\
\\
{\sc Caesar P. Palisoc}\\
{\normalsize National Institute of Physics, University of the Philippines,}\\
{\normalsize Diliman, Quezon City 1101, Philippines}}

\date{}

\maketitle

\begin{abstract}
We study the production of a prompt $J/\psi$ meson in association with a
prompt photon in $ep$ deep-inelastic scattering within the factorisation
formalism of non-relativistic quantum chromodynamics (NRQCD) and demonstrate
that this process provides a clean probe of the colour-octet mechanism at DESY
HERA.
Our analysis is based on an updated set of non-perturbative NRQCD matrix
elements obtained through a joint fit to data on charmonium inclusive
hadroproduction from runs I and II at the Fermilab Tevatron.

\medskip

\end{abstract}

\newpage

\section{Introduction}
\label{intro}

Since the discovery of the $J/\psi$ meson in 1974, charmonium has provided a
useful laboratory for quantitative tests of quantum chromodynamics (QCD) and,
in particular, of the interplay of perturbative and non-perturbative phenomena.
The factorisation formalism \cite{bbl} of nonrelativistic QCD (NRQCD)
\cite{cas} provides a rigorous theoretical framework for the description of
heavy-quarkonium production and decay.
This formalism implies a separation of short-distance coefficients, which can 
be calculated perturbatively as expansions in the strong-coupling constant
$\alpha_s$, from long-distance matrix elements (MEs), which must be extracted
from experiment.
The relative importance of the latter can be estimated by means of velocity
scaling rules; {\it i.e.}, the MEs are predicted to scale with a definite
power of the heavy-quark ($Q$) velocity $v$ in the limit $v\ll1$.
In this way, the theoretical predictions are organised as double expansions in
$\alpha_s$ and $v$.
A crucial feature of this formalism is that it takes into account the complete
structure of the $Q\overline{Q}$ Fock space, which is spanned by the states
$n={}^{2S+1}L_J^{(a)}$ with definite spin $S$, orbital angular momentum
$L$, total angular momentum $J$, and colour multiplicity $a=1,8$.
The hierarchy of the MEs predicted by the velocity scaling rules is explained
for the $J/\psi$, $\psi^\prime$, and $\chi_{cJ}$ mesons in Table~\ref{tab:1}.
In particular, this formalism predicts the existence of colour-octet (CO)
processes in nature.
This means that $Q\overline{Q}$ pairs are produced at short distances in
CO states and subsequently evolve into physical, colour-singlet (CS) quarkonia
by the non-perturbative emission of soft gluons.
In the limit $v\to0$, the traditional CS model (CSM) \cite{ber} is recovered
in the case of $S$-wave quarkonia.
The greatest triumph of this formalism was that it was able to correctly 
describe \cite{bra,cho} the cross section of inclusive charmonium
hadroproduction measured in $p\overline{p}$ collisions at the Fermilab
Tevatron \cite{runi}, which had turned out to be more than one order of
magnitude in excess of the theoretical prediction based on the CSM.
Apart from this phenomenological drawback, the CSM also suffers from severe
conceptual problems indicating that it is incomplete.
These include the presence of logarithmic infrared singularities in the
${\cal O}(\alpha_s)$ corrections to $P$-wave decays to light hadrons and in
the relativistic corrections to $S$-wave annihilation \cite{bar}, and the lack
of a general argument for its validity in higher orders of perturbation
theory.

In order to convincingly establish the phenomenological significance of the
CO processes, it is indispensable to identify them in other kinds of
high-energy experiments as well.
Studies of charmonium production in $ep$ photoproduction, $ep$ and $\nu N$
deep-inelastic scattering (DIS), $e^+e^-$ annihilation in the continuum,
$Z$-boson decays, $\gamma\gamma$ collisions, and $b$-hadron decays may be
found in the literature; for reviews, see \cite{yua}.
Furthermore, the polarisation of $\psi^\prime$ mesons produced directly and
of $J/\psi$ mesons produced promptly, {\it i.e.}, either directly or via the
feed-down from heavier charmonia, which also provides a sensitive probe of CO
processes, was investigated \cite{bkl,lee}.
Until recently, none of these studies was able to prove or disprove the NRQCD
factorisation hypothesis.
However, H1 data of $e+p\to e+J/\psi+X$ in DIS at the DESY Hadron Electron Ring
Accelerator (HERA) \cite{h1,zeus} and DELPHI data of
$\gamma+\gamma\to J/\psi+X$ at the CERN Large Electron Positron Collider (LEP2)
\cite{delphi} provide first independent evidence for it by agreeing with the
respective NRQCD predictions \cite{ep,gg}.

\begin{table}
\caption{Values of $k$ in $\left\langle{\cal O}^H[n]\right\rangle\propto v^k$
for $H=J/\psi,\psi^\prime,\chi_{cJ}$.}
\label{tab:1}
\begin{tabular}{lll}
\hline\noalign{\smallskip}
$k$ & $J/\psi$, $\psi^\prime$ & $\chi_{cJ}$ \\
\noalign{\smallskip}\hline\noalign{\smallskip}
3 & ${}^3\!S_1^{(1)}$ & --- \\
5 & --- & ${}^3\!P_J^{(1)}$, ${}^3\!S_1^{(8)}$ \\
7 & ${}^1\!S_0^{(8)}$, ${}^3\!S_1^{(8)}$, ${}^3\!P_J^{(8)}$ & --- \\
\noalign{\smallskip}\hline
\end{tabular}
\end{table}

In this paper, we identify the DIS process
\begin{equation}
e+p\to J/\psi+\gamma+X
\label{eq:had}
\end{equation}
as a clean probe of the CO mechanism and propose its experimental study at
HERA~II.
In fact, among the partonic subprocesses contributing at LO,
\begin{eqnarray}
e+\gamma&\to&e+c\overline{c}\left[{}^3\!S_1^{(1)}\right]+\gamma,
\label{eq:p}\\
e+g&\to&e+c\overline{c}\left[{}^3\!S_1^{(8)}\right]+\gamma,
\label{eq:g}
\end{eqnarray}
the latter is by far dominant because, in the relevant range of $x$ and $Q^2$,
the density of gluons in the proton is so much higher than the one of photons
that the $O(v^4)$ suppression of
$\left\langle{\cal O}^{J/\psi}\left[{}^3\!S_1^{(8)}\right]\right\rangle$
relative to
$\left\langle{\cal O}^{J/\psi}\left[{}^3\!S_1^{(1)}\right]\right\rangle$
(see Table~\ref{tab:1}) is inconsequential.
The emission of photons off the proton can happen either elastically or
inelastically, {\it i.e.}, the proton stays intact or is destroyed.
In both cases, the parton density function (PDF) can be evaluated in the
Weizs\"acker-Williams approximation \cite{kni,gsv}.
Besides electromagnetic proton interaction, also diffractive scattering off
the proton, via pomeron exchange, allows for CS processes.
However, such events will be accumulated at the border of the phase space, at
$z\alt1$, where $z$ is the inelasticity variable defined in
Sect.~\ref{sec:2}.
By the same token, they can be eliminated from the experimental data set by
applying an appropriate acceptance cut on $z$.
In this paper, we assume this to be done.

The potential of $J/\psi$ plus photon associated production to probe the CO
mechanism was already investigated for photoproduction in $ep$ scattering
\cite{cac}.
In that case, however, the bulk of the cross section, more than 2/3 for
$J/\psi$ transverse momenta $p_T>1$~GeV (see Table~II of \cite{cac}), is
due to the CS channel
$g+g\to c\overline{c}\left[{}^3\!S_1^{(1)}\right]+\gamma$ in resolved
photoproduction.
Specifically, the photoproduction analogue of process~(\ref{eq:g}),
$\gamma+g\to c\overline{c}\left[{}^3\!S_1^{(8)}\right]+\gamma$, only makes up
1/5 of the total cross section \cite{cac}.
On the other hand, the probability of a photon to appear resolved rapidly
decreases with its size $1/Q^2$, so that the situation encountered in
\cite{cac} is subject to a dramatic change as one passes from
photoproduction to DIS.

In \cite{nlog}, prompt $J/\psi$ plus photon associated photoproduction in
$\gamma\gamma$ collisions was studied in next-to-leading order (NLO), and
sizeable corrections to the cross section were found.
Since the present analysis is at an exploratory level, considering
process~(\ref{eq:had}) for the first time, we stay at LO.
A NLO analysis of process~(\ref{eq:had}) would be more involved than in the
case of \cite{nlog}, due to the presence of $Q^2$ as an additional mass
scale, and is left for future work.

The paper is organised as follows.
In Sect.~\ref{sec:2}, we collect the formulas from which the LO cross
section of process~(\ref{eq:had}) can be evaluated.
In Sect.~\ref{sec:3}, we update the extraction of the NRQCD MEs of the
$J/\psi$, $\psi^\prime$, and $\chi_{cJ}$ mesons in \cite{bkl} by
including in the fit CDF data from Tevatron run~II \cite{runii} besides that
from run~I \cite{runi}. 
In Sect.~\ref{sec:4}, we then present our predictions for the cross
section of process~(\ref{eq:had}) under HERA~II experimental conditions and
demonstrate that this is an excellent probe of the CO mechanism.
Our conclusions are contained in Sect.~\ref{sec:5}.

\section{Analytic results}
\label{sec:2}

We now present our analytic results for the cross section of
process~(\ref{eq:had}).
We work at LO in the parton model of QCD with $n_f=3$ active quark flavours
and employ the NRQCD factorisation formalism \cite{bbl} to describe the
formation of the $J/\psi$ meson.
We start by defining the kinematics.
We denote the four-momenta of the incoming lepton and proton and the outgoing
lepton, $J/\psi$ meson, and photon by $k$, $P$, $k^\prime$, $p_\psi$, and
$p^\prime$, respectively.
The parton $a$ struck by the virtual photon ($\gamma^\star$) carries
four-momentum $p=xP$.
The virtual photon has four-momentum $q=k-k^\prime$, and it is customary to
define $Q^2=-q^2>0$, $y=q\cdot P/k\cdot P$, and $z=p_\psi\cdot P/q\cdot P$.
In the proton rest frame, $y$ and $z$ measure the relative lepton energy loss
and the fraction of the virtual-photon energy transferred to the $J/\psi$
meson, respectively.
We neglect the masses of the proton, lepton, and light quarks, call the one
of the $J/\psi$ meson $M_\psi$, and take the charm-quark mass to be
$m_c=M_\psi/2$.
In our approximation, the proton remnant $X$ has zero invariant mass,
$M_X^2=(P-p)^2=0$.
The centre-of-mass (c.m.) energy squares of the $ep$ and $\gamma^\star p$
collision are $S=(k+P)^2$ and $W^2=(q+P)^2=yS-Q^2$, respectively.
As usual, we define the partonic Mandelstam variables as
$s=(q+p)^2=xyS-Q^2$, $t=(q-p_\psi)^2=-xy(1-z)S$, and
$u=(p-p_\psi)^2=M_\psi^2-xyzS$.
By four-momentum conservation, we have $s+t+u=M_\psi^2-Q^2$.
In the $\gamma^\star p$ CM frame, the $J/\psi$ meson has transverse momentum
and rapidity
\begin{eqnarray}
p_T^\star&=&\frac{\sqrt{t\left(su+Q^2M_\psi^2\right)}}{s+Q^2},
\\
y_\psi^\star&=&\frac{1}{2}\ln\frac{s\left(M_\psi^2-u\right)}
{s\left(M_\psi^2-t\right)+Q^2M_\psi^2}+\frac{1}{2}\ln\frac{W^2}{s},
\label{eq:ycms}
\end{eqnarray}
respectively.
Here and in the following, we denote the quantities referring to the
$\gamma^\star p$ CM frame by an asterisk.
The second term on the right-hand side of (\ref{eq:ycms}) originates from 
the Lorentz boost from the $\gamma^\star a$ CM frame to the $\gamma^\star p$
one.
Here, $y_\psi^\star$ is taken to be positive in the direction of the
three-momentum of the virtual photon, in accordance with HERA conventions
\cite{h1,zeus}.

The cross sections of processes~(\ref{eq:p}) and (\ref{eq:g}) may be
conveniently calculated by applying the covariant-projector method of
\cite{pet}.
They are related to the one of
$e+g\to e+c\overline{c}\left[{}^3\!S_1^{(8)}\right]+g$, given in (13) of
\cite{ep}, by 
\begin{eqnarray}
\lefteqn{\frac{d^3\sigma}{dy\,dQ^2\,dt}
\left(e+\gamma\to e+c\overline{c}\left[{}^3\!S_1^{(1)}\right]+\gamma\right)}
\nonumber\\
&=&\frac{64}{9}\,\frac{\alpha^2}{\alpha_s^2}\,\frac{d^3\sigma}{dy\,dQ^2\,dt}
\left(e+g\to e+c\overline{c}\left[{}^3\!S_1^{(1)}\right]+g\right),
\nonumber\\
\lefteqn{\frac{d^3\sigma}{dy\,dQ^2\,dt}
\left(e+g\to e+c\overline{c}\left[{}^3\!S_1^{(8)}\right]+\gamma\right)}
\nonumber\\
&=&2\frac{\alpha}{\alpha_s}\,\frac{d^3\sigma}{dy\,dQ^2\,dt}
\left(e+g\to e+c\overline{c}\left[{}^3\!S_1^{(1)}\right]+g\right),
\label{eq:xs}
\end{eqnarray}
where the proportionality factors account for colour and coupling adjustments.

According to the factorisation theorems of the parton model and NRQCD, the
cross section of process (\ref{eq:had}) is then evaluated as
\begin{eqnarray}
\lefteqn{\frac{d^2\sigma}{dy\,dQ^2}(e+p\to e+J/\psi+\gamma+X)}
\nonumber\\
&=&\int_{(Q^2+M_\psi^2)/(yS)}^1dx\int_{-(s+Q^2)(s-M_\psi^2)/s}^0dt
\nonumber\\
&&{}\times
\sum_af_{a/p}(x,M)\sum_n\left\langle{\cal O}^{J/\psi}[n]\right\rangle
\nonumber\\
&&{}\times
\frac{d^3\sigma}{dy\,dQ^2\,dt}(e+a\to e+c\overline{c}[n]+\gamma),
\label{eq:dif}
\end{eqnarray}
where the sums run over $(a,n)=\left(\gamma,{}^3\!S_1^{(1)}\right),
\left(g,{}^3\!S_1^{(8)}\right)$, $f_{a/p}(x,M)$ is the PDF of parton $a$ in
the proton at factorisation scale $M$, and
$\left(d^3\sigma/dy\,dQ^2\,dt\right)(e+a\to e+c\overline{c}[n]+\gamma)$
are given by (\ref{eq:xs}).
The kinematically allowed ranges of $y$ and $Q^2$ are $M_\psi^2/S<y<1$ and
$0<Q^2<yS-M_\psi^2$, respectively.

Prompt $J/\psi$ production may be conveniently described by inserting in
(\ref{eq:dif}) the effective MEs specified in (20) of
\cite{nlog}.

\section{Determination of the MEs}
\label{sec:3}

The recent CDF measurement of prompt $J/\psi$ inclusive hadroproduction in
run~II at the Tevatron (with $\sqrt S=1.96$~TeV and $|y_\psi|<0.6$)
\cite{runii} allows us to update and improve our knowledge of the CO MEs.
Previous determinations by one of us in collaboration with Kramer \cite{kk}
and with Braaten and Lee \cite{bkl} were only based on run~I data
(with $\sqrt S=1.8$~TeV and $|y_\psi|<0.6$) \cite{runi}.
However, the latter data were more detailed because the prompt $J/\psi$ sample
was explicitly broken down into direct $J/\psi$ mesons, feed-down from
$\psi^\prime$ mesons, and feed-down from $\chi_{cJ}$ mesons.
In order to make maximum use of the available information, we perform a joint
fit to the data from runs I and II, which otherwise proceeds along the lines of
\cite{bkl}.
First, the CS MEs
$\left\langle{\cal O}^{\psi(nS)}\left[{}^3\!S_1^{(1)}\right]\right\rangle$,
with $n=1,2$, and
$\left\langle{\cal O}^{\chi_{c0}}\left[{}^3\!P_0^{(1)}\right]\right\rangle$
are extracted from the measured partial decay widths of $\psi(nS)\to l^++l^-$
and $\chi_{c2}\to\gamma+\gamma$ \cite{pdg}, respectively.
Then, the CO MEs
$\left\langle{\cal O}^{\psi(nS)}\left[{}^3\!S_1^{(8)}\right]\right\rangle$,
$\left\langle{\cal O}^{\psi(nS)}\left[{}^1\!S_0^{(8)}\right]\right\rangle$,
$\left\langle{\cal O}^{\psi(nS)}\left[{}^3\!P_0^{(8)}\right]\right\rangle$,
and
$\left\langle{\cal O}^{\chi_{c0}}\left[{}^3\!S_1^{(8)}\right]\right\rangle$
are fitted to the $p_T$ distributions of $\psi(nS)$ and $\chi_{cJ}$ inclusive
hadroproduction \cite{runi,runii} and the cross-section ratio
$\sigma_{\chi_{c2}}/\sigma_{\chi_{c1}}$ \cite{aff} measured at the Tevatron.
In contrast to the run~I data \cite{runi}, the run~II data \cite{runii} reach
down to $p_T=0$.
It turns out that the run~II data in the newly covered low-$p_T$ range are
comparable to or below the CSM prediction, so that their inclusion would spoil
the fit.
Therefore, we only include the 14 data points with $p_T>4.25$~GeV. 
Since the fit results for
$\left\langle{\cal O}^{\psi(nS)}\left[{}^1\!S_0^{(8)}\right]\right\rangle$ and
$\left\langle{\cal O}^{\psi(nS)}\left[{}^3\!P_0^{(8)}\right]\right\rangle$ are
strongly correlated, we consider the linear combination
\begin{equation}
M_r^{\psi(nS)}
=\left\langle{\cal O}^{\psi(nS)}\left[{}^1\!S_0^{(8)}\right]\right\rangle
+\frac{r}{m_c^2}
\left\langle{\cal O}^{\psi(nS)}\left[{}^3\!P_0^{(8)}\right]\right\rangle,
\label{eq:mr}
\end{equation}
where the value of $r$ is chosen so that the error on $M_r^{\psi(nS)}$ is
minimised.
The minimisation in $\chi^2$ is performed exactly, by solving a set of seven
linear equations for the seven unknown CO MEs.

The relevant partonic cross sections may be found in \cite{cho}.
We take the charm-quark mass to be $m_c=(1.5\pm0.1)$~GeV and adopt the
relevant feed-down and leptonic decay branching fractions from
\cite{pdg}.
As for the proton PDFs, we use the latest LO sets by Martin, Roberts,
Stirling, and Thorne (MRST2001LO) \cite{mrst} and the Coordinated
Theoretical-Experimental Project on QCD\break (CTEQ6L1) \cite{cteq}.
For consistency, we employ the one-loop formula for $\alpha_s^{(3)}(\mu)$ and
choose the asymptotic scale parameter to be $\Lambda_{\rm QCD}^{(3)}=253$~MeV
(246~MeV), appropriate for the MRST2001LO (CTEQ6L1) PDFs.
We identify the renormalisation scale $\mu$ and the factorisation scale $M$
with the charmonium transverse mass $m_T=\sqrt{p_T^2+M_\psi^2}$.

The sets of MEs thus obtained with the MRST2001LO and CTEQ6L1 PDFs are
summarised in Table~\ref{tab:2}.
The respective values of $\chi^2$ per degree of freedom are 49/52 and 50/52.
The quoted errors are of experimental origin only.
Comparing Table~\ref{tab:2} with Table~I of \cite{bkl}, where previous
MRST and CTEQ PDFs were used, we observe that the MEs obtained with CTEQ PDFs
are only moderately changed.
On the other hand, the MRST values of
$\left\langle{\cal O}^{\psi(nS)}\left[{}^3\!S_1^{(8)}\right]\right\rangle$ and
$\left\langle{\cal O}^{\chi_{c0}}\left[{}^3\!S_1^{(8)}\right]\right\rangle$ in
Table~\ref{tab:2} are appreciably smaller than in \cite{bkl}.

\begin{table*}
\caption{NRQCD MEs of the $J/\psi$, $\psi^\prime$, and $\chi_{cJ}$ obtained as
described in the text using the MRST2001LO \cite{mrst} and CTEQ6L1 \cite{cteq}
PDFs.
The errors are experimental only}
\label{tab:2}
%{\tiny
\begin{tabular}{lll}
\hline\noalign{\smallskip}
& MRST2001LO \cite{mrst} & CTEQ6L1 \cite{cteq} \\
\noalign{\smallskip}\hline\noalign{\smallskip}
$\left\langle{\cal O}^{J/\psi}\left[{}^3\!S_1^{(1)}\right]\right\rangle$ &
$1.5\pm 0.1$ GeV${}^3$ &
$1.4\pm 0.1$ GeV${}^3$ \\
$\left\langle{\cal O}^{J/\psi}\left[{}^3\!S_1^{(8)}\right]\right\rangle$ &
$(3.9\pm 1.4)\times 10^{-4}$ GeV${}^3$ &
$(2.3\pm 0.2)\times 10^{-3}$ GeV${}^3$ \\
$M_{3.7,3.6}^{J/\psi}$ &
$(6.0\pm 0.1)\times 10^{-2}$ GeV${}^3$ &
$(7.3\pm 0.2)\times 10^{-2}$ GeV${}^3$ \\
$\left\langle{\cal O}^{\psi^\prime}\left[{}^3\!S_1^{(1)}\right]\right\rangle$ &
$(6.7\pm 0.5)\times 10^{-1}$ GeV${}^3$ & 
$(6.7\pm 0.5)\times 10^{-1}$ GeV${}^3$ \\ 
$\left\langle{\cal O}^{\psi^\prime}\left[{}^3\!S_1^{(8)}\right]\right\rangle$ &
$(8.8\pm 1.2)\times 10^{-4}$ GeV${}^3$ & 
$(2.0\pm 0.2)\times 10^{-3}$ GeV${}^3$ \\
$M_{3.5}^{\psi^\prime}$ &
$(1.1\pm 0.1)\times 10^{-2}$ GeV${}^3$ &
$(1.0\pm 0.2)\times 10^{-2}$ GeV${}^3$ \\
$\left\langle{\cal O}^{\chi_{c0}}\left[{}^3\!P_0^{(1)}\right]\right\rangle$ &
$(1.2\pm 0.1)\times 10^{-1}$ GeV${}^5$ & 
$(1.2\pm 0.1)\times 10^{-1}$ GeV${}^5$ \\ 
$\left\langle{\cal O}^{\chi_{c0}}\left[{}^3\!S_1^{(8)}\right]\right\rangle$ &
$(4.7\pm 0.6)\times 10^{-4}$ GeV${}^3$ & 
$(1.1\pm 0.1)\times 10^{-3}$ GeV${}^3$ \\
\noalign{\smallskip}\hline
\end{tabular}
%}
\end{table*}

\boldmath
\section{Prompt $J/\psi$ plus photon associated electroproduction}
\label{sec:4}
\unboldmath

We are now in a position to present our theoretical predictions for the cross
section of process~(\ref{eq:had}) under HERA~II experimental conditions.
They are evaluated from the formulas listed in Sect.~\ref{sec:2} with the
inputs and conventions described in Sect.~\ref{sec:3}, except that we now set
$\mu=M=\xi\sqrt{Q^2+M^2}$ and vary the scale parameter $\xi$ between 1/2 and 2
about the default value 1.
Since (\ref{eq:dif}) is sensitive to a different linear combination of
$\left\langle{\cal O}^{\psi(nS)}\left[{}^1\!S_0^{(8)}\right]\right\rangle$ and
$\left\langle{\cal O}^{\psi(nS)}\left[{}^3\!P_0^{(8)}\right]\right\rangle$
than the one appearing in (\ref{eq:mr}), we write
\begin{eqnarray}
\left\langle{\cal O}^{\psi(nS)}\left[{}^1\!S_0^{(8)}\right]\right\rangle
&=&\kappa M_r^{\psi(nS)},
\nonumber\\
\left\langle{\cal O}^{\psi(nS)}\left[{}^3\!P_0^{(8)}\right]\right\rangle
&=&(1-\kappa)\frac{m_c^2}{r}M_r^{\psi(nS)}
\end{eqnarray}
and vary $\kappa$ between 0 and 1 around the default value 1/2.

In order to estimate the theoretical uncertainties in our predictions, we
vary the unphysical parameters $\xi$ and $\kappa$ as indicated above, take
into account the experimental errors on $m_c$ and the default MEs, and switch
from the MRST2001LO PDFs \cite{mrst}, which we take as our default, to the
CTEQ6L1 ones \cite{cteq}, properly adjusting $\Lambda_{\rm QCD}^{(3)}$ and the
MEs.
We then combine the individual shifts in quadrature, allowing for the upper
and lower half errors to be different.

In the laboratory frame of HERA~II, electrons or posi\-trons with energy
$E_e=27.5$~GeV collide with protons of energy $E_p=920$~GeV, yielding a c.m.\
energy of $\sqrt{S}=2\sqrt{E_eE_p}=318$~GeV.
For definiteness, we adopt the experimental acceptance cuts from the recent
H1 analysis of inclusive $J/\psi$ electroproduction \cite{h1}, which include
$2<Q^2<100$~GeV, $50<W<225$~GeV, $0.3<z<0.9$, and $p_T^{\star2}>1$~GeV${}^2$.
We consider cross section distributions in $Q^2$, $W$, $z$, $p_T^{\star2}$,
$y_\psi^\star$, $p_T^2$, and $y_\psi$, where the last four variables refer to
the $J/\psi$ meson, adopting the binning from \cite{h1}.

\begin{figure*}
\resizebox{0.75\textwidth}{!}{%
  \includegraphics{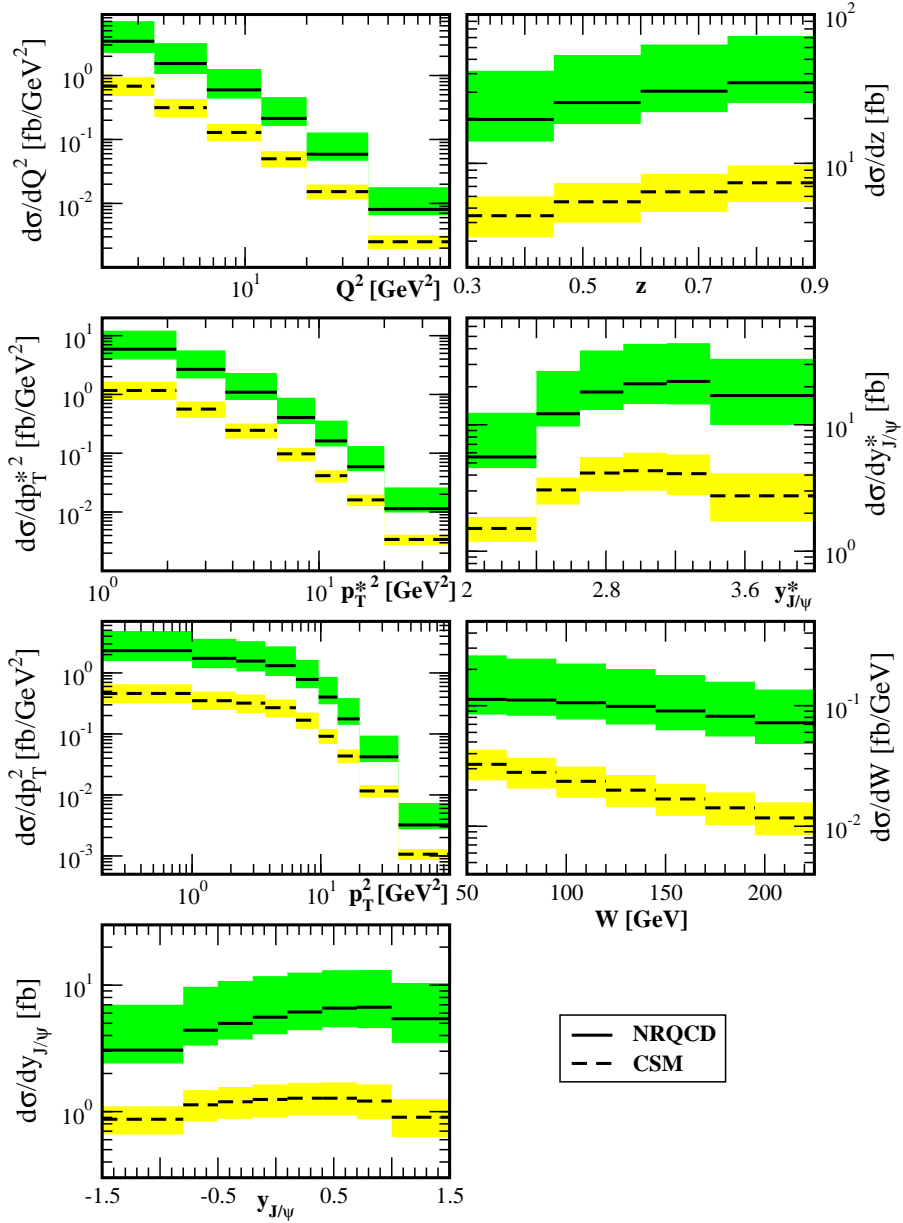}
}
\caption{NRQCD (solid lines) and CSM (dashed lines) predictions of the $Q^2$,
$z$, $p_T^{\star2}$, $y_\psi^\star$, $p_T^2$, $W$, and $y_\psi$ distributions
of prompt $J/\psi$ plus photon associated electroproduction at HERA~II in the
kinematic region defined by $2<Q^2<100$~GeV, $50<W<225$~GeV, $0.3<z<0.9$, and
$p_T^{\star2}>1$~GeV${}^2$.
The shaded bands indicate the theoretical uncertainty}
\label{fig:1}
\end{figure*}

Our results are displayed in Fig.~\ref{fig:1}, where the NRQCD predictions
(solid lines) are compared with the CSM ones (dashed lines).
In each case, the theoretical uncertainties are indicated by the shaded bands.
As expected from our discussion in Sect.~\ref{intro}, the NRQCD
predictions vastly exceed the CSM ones, by almost one order of magnitude,
The gaps are considerably larger than the theoretical errors.
On the other hand, the shapes of the various cross section distributions
come out very similar in both approaches, which may be understood by observing
that the partonic cross sections of the CS and CO channels in
(\ref{eq:xs}) are proportional to each other.
From Fig.~\ref{fig:1}, we read off that the integrated NRQCD cross section is
of order 10~fb.
Given that the integrated luminosity to be collected by the end of HERA
operation amounts to about 1~fb${}^{-1}$, one thus expects about 10 signal
events.

\section{Conclusions}
\label{sec:5}

We studied the electroproduction of prompt $J/\psi$ mesons in association with
prompt photons in $ep$ collisions under HERA~II kinematic conditions to LO in
the NRQCD \cite{cas} factorisation formalism \cite{bbl}.
We considered cross section distributions in all variables of current interest
\cite{h1,zeus}, including $Q^2$, $W$, $z$, $p_T^{\star2}$, $y_\psi^\star$,
$p_T^2$, and $y_\psi$.
As input for our calculation, we used updated information on the NRQCD MEs
extracted from a combined fit to data on inclusive charmonium hadroproduction
collected by the CDF Collaboration in runs~I \cite{runi} and II \cite{runii}
at the Tevatron.

As a result of our study, we could identify prompt $J/\psi$ plus photon
associated electroproduction as a useful probe of the CO mechanism.
In fact the NRQCD predictions turned out to exceed the CSM ones by almost one
order of magnitude.
Unfortunately, the cross section distributions in both theories have very
similar shapes, so that the NRQCD to CSM ratio cannot be further enhanced by
specific acceptance cuts.
Should this production process be experimentally observed with the rate
predicted by NRQCD, then this would provide strong evidence in favour of the
existence of CO processes in nature.
In view of the moderate integrated cross section of order 10~fb, this is a
challenging endeavour.
This is an example of a study that would benefit from the extension of HERA
operation beyond the summer of 2007.

\vspace{0.5cm}

\noindent
{\it Acknowledgements.}
The research of B.A.K. was supported in part by the BMBF through Grant No.\
05~HT6GUA and the DFG through Grant No.\ KN~365/6--1.
The research of C.P.P. was supported in part by the DFG through
Graduiertenkolleg No.\ GRK~602/2 and by the Office of the Vice President for
Academic Affairs of the University of the Philippines.

\end{document}